\documentclass[english,apl,reprint]{revtex4-1}
\usepackage[LGR,T1]{fontenc}
\usepackage{textcomp}
\usepackage{amsmath}
\usepackage{graphicx}

\makeatletter

\DeclareRobustCommand{\greektext}{%
  \fontencoding{LGR}\selectfont\def\encodingdefault{LGR}}
\DeclareRobustCommand{\textgreek}[1]{\leavevmode{\greektext #1}}
\ProvideTextCommand{\~}{LGR}[1]{\char126#1}

\makeatother

\usepackage{babel}
\begin{document}

\title{High Quality Factor Surface Fabry-Perot Cavity of Acoustic Waves}

\author{Yuntao Xu}

\affiliation{Department of Electrical Engineering, Yale University, New Haven,
CT 06511, USA}

\author{Wei Fu}

\affiliation{Department of Electrical Engineering, Yale University, New Haven,
CT 06511, USA}

\author{Chang-ling Zou}

\affiliation{Department of Electrical Engineering, Yale University, New Haven,
CT 06511, USA}

\author{Zhen Shen}

\affiliation{Department of Electrical Engineering, Yale University, New Haven,
CT 06511, USA}

\author{Hong X. Tang}
\email{hong.tang@yale.edu}

\affiliation{Department of Electrical Engineering, Yale University, New Haven,
CT 06511, USA}

\date{\today}
\begin{abstract}
Surface acoustic wave (SAW) resonators are critical components in
wireless communications and many sensing applications. They have also
recently emerged as subject of study in quantum acoustics at the single
phonon level. Acoustic loss reduction and mode confinement are key
performance factors in SAW resonators. Here we report the design and
experimental realization of a high quality factor Fabry-Perot SAW
resonators formed in between tapered phononic crystal mirrors patterned
on a GaN-on-sapphire material platform . The fabricated SAW resonators
are characterized by both electrical network analyzer and optical
heterodyne vibrometer. We observed standing Rayleigh wave inside the
cavity, with an intrinsic quality factor exceeding $1.3\times10^{4}$
at ambient conditions. 
\end{abstract}
\maketitle

\section{Introduction}

In last decades, surface acoustic wave (SAW) devices have found wide
use in analog signal processing \cite{morgan2010surface,campbell2012surface},
wireless communications \cite{campbell1998surface,hashimoto2000surface}
and a range of sensing applications \cite{wohltjen1984mechanism,lange2008surface}.
Recently, SAW based quantum acoustics have received considerable attention
for their flexibility in coupling with various quantum systems, including
the superconducting qubits \cite{Chueaao1511,gustafsson2014propagating},
NV centers in diamond \cite{golter2016coupling}, quantum dots \cite{schuetz2017acoustic},
and is potential for building hybrid quantum networks \cite{Schutz2017}.
Despite their significance in classical and quantum information processing,
achieving low loss resonator and efficient mode confinement are still
challenging. 

By analogy to optics, an acoustic resonator can be constructed in
the form of Fabry-Perot (FP) cavity in which the SAW is bounced between
two mirrors and confined in between. Unlike optics where highly reflective
mirrors can be patterned from a high-refractive-index waveguiding
layer, unsuspended, directly etched acoustic reflectors are often
insufficient in producing high enough reflectivity to prevent coupling
to bulk acoustic modes \cite{staples1974uhf}. By shorting interdigital
transducers (IDT), the SAW can be confined more gently because of
the small velocity contrast between the region with and without metal
cladding \cite{chen1985analysis}. This gentle confinement however
requires a large number of IDTs and hence very large device area for
high reflectivity which in turn could cause SAW suffering from the
Ohmic loss of the metal, leading to limited quality factor. Recently,
such FP type SAW resonators have been demonstrated in the context
of quantum acoustics utilizing superocnducting electrodes, with Q
reaching 0.5 million but the devices are limited to operate only at
cryogenic temperatures \cite{manenti2016surface}. 

Recently it was shown that it is feasible to confine SAW with phononic
band gap structures\cite{liu2014design,wang2015tapered}. Challenges
remain in achieving full acoustic confinement and bringing the quality
factor of such acoustic resonators to the value of their suspended
counter part, which is fragile and requires micromachining processing.
Here we report a SAW FP cavity formed by surface patterned phononic
crystal (PnC) mirrors which provide high reflectivity with a much
reduced mirror length, thus enabling high-Q room temperature SAW resonators.
By harnessing the engineered band gap of phononic structure \cite{khelif2015phononic,laude2015phononic,maldovan2015phonon,zen2014engineering,benchabane2015guidance},
we design the cavity for the Rayleigh surface modes. The device is
fabricated by GaN-on-sapphire material system, and the cavity modes
are excited by intracavity IDTs. The microwave reflection spectrum
shows very high Q resonances with predicted free spectral range (FSR)
, indicating the strong confinement of SAW modes and enhanced coupling
to the IDTs. In addition, the cavity acoustic field is also probed
by a custom-built optical vibrometer, allowing for the identification
of the confined SAW modes. The demonstrated devices pave a new direction
for the study of strong phonon-matter interaction as an analogue to
the cavity quantum electrodynamics \cite{soykal2011sound} and the
cavity optomechanics \cite{okada2017cavity}. The demonstrated high
Q resonators are also promising for sensing applications based on
surface acoustic wave devices.

\section{Design}

The acoustic FP resonators are patterned from Gallium Nitride (GaN)
(0001) epitaxially grown on c-plane sapphire substrate \cite{bruch2015broadband}.
GaN is a piezoelectric material with low mechanical damping \cite{rais2014gallium}.
Due to its lower acoustic shear velocity than that of the sapphire
substrate, the acoustic waves can be guided by the GaN layer, where
an acoustic cavity can be further formed by lateral confinement through
a pair of PnC acoustic mirrors, schematically shown in Fig. 1(a).
IDTs are directly placed within the cavity atop the GaN layer to excite
SAW. At cavity resonances, standing waves are built-up within the
excitation bandwidth of the IDTs. Within the cavity, the coupling
between acoustic waves and the IDTs are enhanced and therefore the
number of IDT electrodes can be minimized (5 pairs are utilized in
a typical device). To avoid the scattering loss to the substrate,
we adapted a tapered mirror design which is first proposed in Ref.
\cite{wang2015tapered} for achieving efficient acoustic wave transitions
in the phononic crystals . 

\begin{figure}
\includegraphics[width=0.445\textwidth]{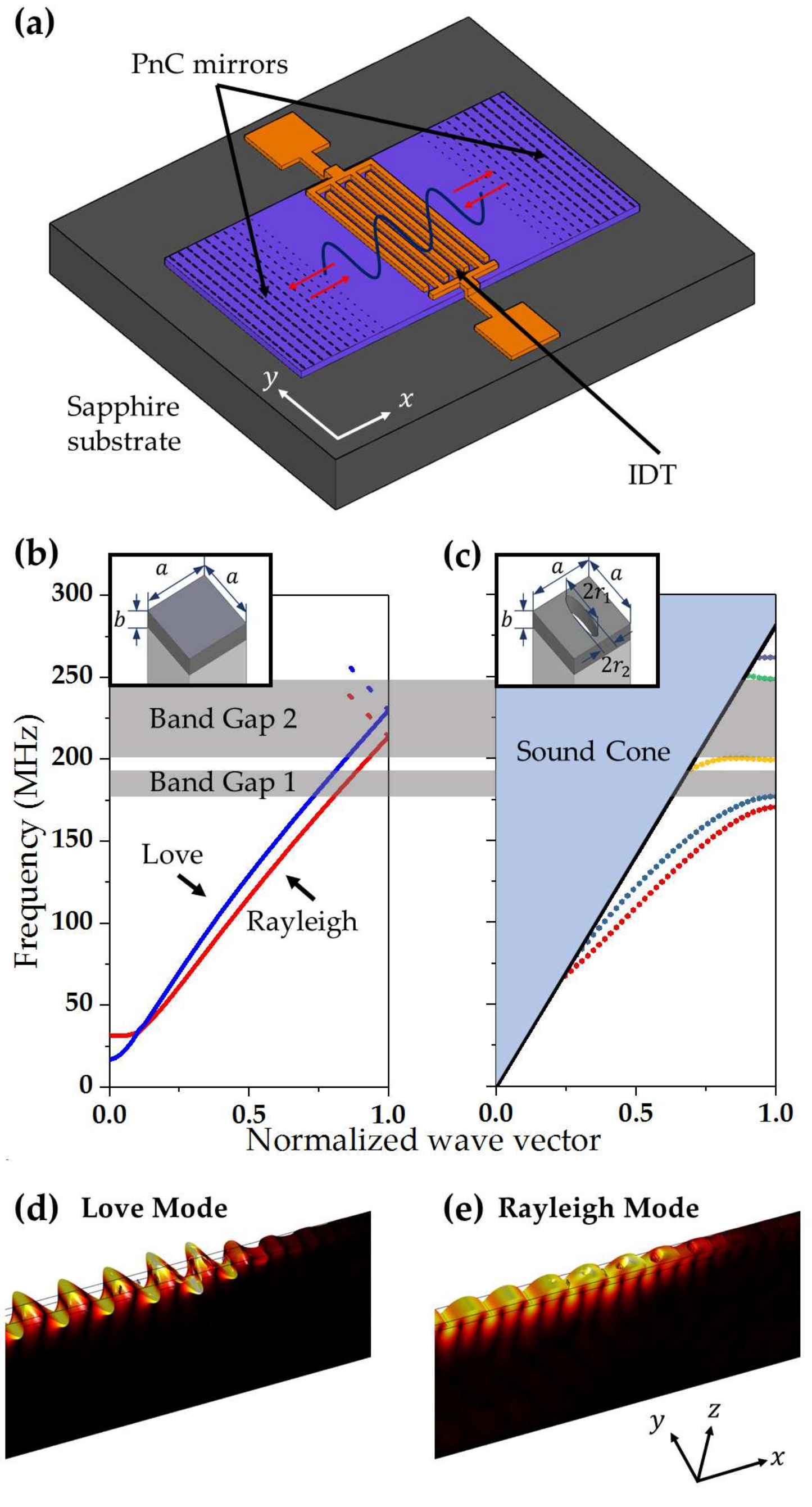}\caption{(a) Schematic illustration of acoustic Fabry-Perot cavity design where
the SAW is confined by a pair of square lattices PnC mirror and excited
by IDTs. (b) The dispersion of Rayleigh and Love mode without PnC.
The unit cell is set to have a period of $a=11\,\mathrm{\mu m}$ and
thickness of $b=5\,\mathrm{\mu m}$. (c) The phononic band structures
of GaN-on-sapphire square PnC with elliptical hole in each unit cell.
The unit cell is shown as inset and $a$, $b$, $r_{1}$, $r_{2}$
represent the period of square lattice, the thickness of GaN , the
major and minor radius of elliptical hole, respectively. Here the
parameters are set as $a=11\,\mathrm{\mu m}$, $b=5\,\mathrm{\mu m}$,
$r_{1}=4.4\,\mathrm{\mu m}$, $r_{2}=0.88\,\mathrm{\mu m}$. Different
colors represent different groups of modes in PnC unit cell: red and
yellow modes have dominant out-of-plane displacement; blue and green
modes have dominant in plane displacement. (d) Simulated displacement
profile of Love mode in the SAW cavity. (e) Simulated displacement
profile of Rayleigh mode in the SAW cavity.}
\end{figure}

The unit cell of the PnC structure, a square pillar with a centered
elliptical hole in GaN layer, is shown in the inset of Fig. 1(c),
where $a$ is the period of the square lattice for both $x$ and $y$
directions, $b$ is the thickness of GaN layer, $r_{1}$, $r_{2}$
are the major and minor radius of the elliptical hole respectively.
Finite element method (FEM) simulation is performed to determine geometrical
parameters of PnC unit cell using published material elastic and piezoelectric
coefficients\cite{pedros2005anisotropy}. By gradully increasing the
aspect ratio of elliptical PnC hole, larger phononic band gap and
better comfinement of SAW can be obtained. The band structure of optimized
PnC unit cell is shown in Fig. 1(c), in which the parameters are chosen
to be $a=11\,\mathrm{\mu m}$, $b=5\,\mathrm{\mu m}$, $r_{1}=4.4\,\mathrm{\mu m}$,
$r_{2}=0.88\,\mathrm{\mu m}$. As a comparison, the band structure
for a unit cell without hole is shown in Figure 1b. The band structures
are obtained by calculating the eigenmode of the PnC unit cell for
different wave vectors along $x$ direction. The lightly shaded area
of the band diagram represent the sound cone, in which region surface
modes may coupled with bulk modes and therefore are considered to
be lossy. Grey shaded areas in Figs. 1(b) and (c) indicate two phonon
band gaps for wave-vectors along $x$ direction, while gap-free Rayleigh
mode and Love mode are supported on flat surface without PnC structure. 

The performance of PnC resonator is then analyzed with FEM simulation.
Due to the relatively large transverse dimension (width) of the cavity,
the resonator could be approximated to have translational symmetry
along transverse ($y$) direction. Hence we only simulate one period
in $y$ direction with periodical boundary condition applied. 40 periods
of elliptical holes are placed on each side of cavity regime with
a length of $l=525\mathrm{\mu m}$. The size of PnC holes is linearly
tapered from $1\,\mathrm{\mu m}$ to $4.4\,\mathrm{\mu m}$ within
7 periods starting from the cavity mirror interface. While the simulation
shows that love mode resonances are supported in both band gap 1 and
band gap 2, Rayleigh mode resonances with good confinement are only
found in band gap 1. At the frequency regime of band gap 2, Rayleigh
wave is scattered by the PnC holes at cavity boundary, which induce
a coupling between SAW and bulk mode in the sound cone, therefore
prevent a strong confinement of surface mode. Hence the frequency
range of band gap 1 (180 MHz to 200MHz) is chosen to design the SAW
FP cavity. The modal profiles of the simulated eigenmodes are given
in Figs. 1(d) and (e), including both Love modes (with mainly in-plane
displacement) and Rayleigh modes (with mainly out-of-plane displacement). 

\section{Fabrication and characterization}

The device fabrication starts from the growth of $5\,\mathrm{\mu m}$
GaN on c-plane sapphire wafers by metal-organic chemical vapour depositon
(MOCVD) \cite{bruch2015broadband}. To fabricate the PnC resonator,
a $1.2\,\mathrm{\mu m}$ layer of $\mathrm{SiO_{2}}$ is deposited
with plasma-enhanced chemical vapor deposition (PECVD). The PnC structures
are patterned with electron beam lithography (EBL) using polymethyl
methacrylate (PMMA) resist, followed by a lift-off process to deposit
a $40\,\mathrm{nm}$ thin layer of chromium. After liftoff, the pattern
are transferred to $\mathrm{SiO_{2}}$ layer by florine-based reactive
ion-etching (RIE), then subsequently transferred to GaN layer by chlorine-based
RIE. A second EBL process is carried on using a bi-layer PMMA resist
to deposit the IDTs by a lift-off process (50nm chromium and 50nm
gold). 

\begin{figure}
\includegraphics[width=0.445\textwidth]{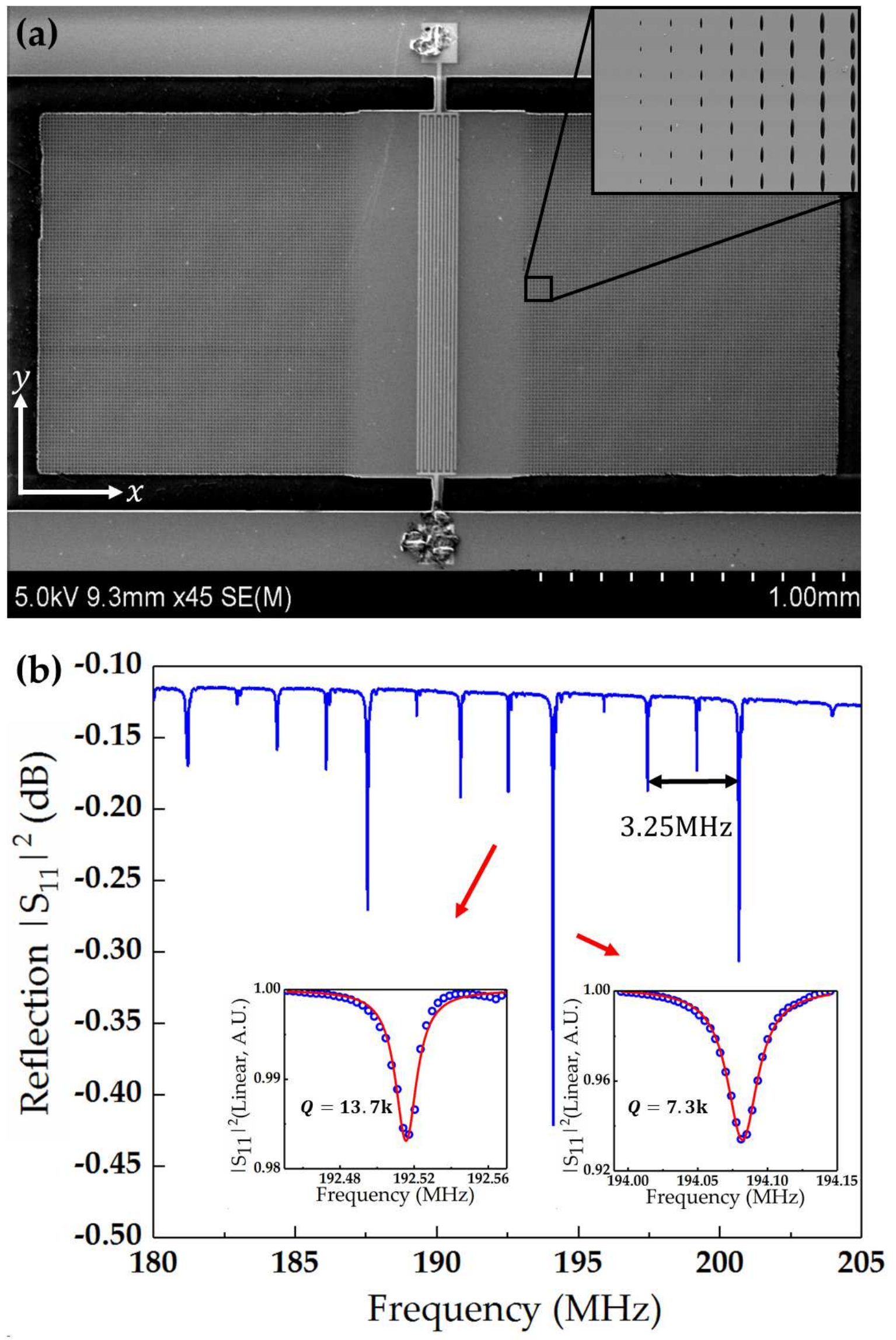}\caption{(a) SEM image of a fabricated PnC SAW resonator (top view) with 100
periods of elliptical holes on each side of the SAW cavity. (b) Microwave
reflection spectrum of fabricated SAW cavity indicates high Q resonances
with a free spectral range (FSR) of 3.25 MHz. The bandwidth of IDTs
could be estimated by $f_{0}/2N\sim20\,\mathrm{MHz}$. Insets show
the zoom in spectrum (symbols) of an acoustic mode with highest quality
factor (left inset) and another acoustic mode with largest extinction
ratio (right inset) fitted by a Lorentz function. }
\end{figure}

A scanning electron microscope (SEM) image of the fabricated device
is shown in Fig.$\,$2(a). The size of cavity regime is $1.1\,\mathrm{mm}$
wide (in $y$ direction) and $0.5\,\mathrm{mm}$ long (in $x$ direction),
with a pair of PnC mirrors on the ends of cavity. Each mirror contains
a square lattice PnC with $100\times100$ elliptical holes whose dimensions
are identical with the simulation. The inset shows the first eight
columns of PnC holes. A chip with several fabricated devices is placed
on a printed circuit board (PCB) with SMA connectors and wire bonded
for the electrical characterization.

The reflection spectrum around the working frequency of the resonator
is recorded with a vector network analyzer. As shown in Fig. 2(b),
a series of Lorentzian-shaped dips is clearly observed, indicating
a group of well confined resonance modes supported in the cavity.
All modes share an approximately same free spectral range (around
$3.25\,\mathrm{MHz}$, as marked in the figure), suggesting that they
have similar group velocities but the modes within one FSR should
have different modal profiles. Based on group veolcity of Rayleigh
mode (the confirmation of mode type is made through vibrometer measurement
and will be discussed later) $v_{R}=4705.8\,\mathrm{m/s}$ from simulation,
the effective cavity length $L=724\,\mathrm{\mu m}$ could be estimated
as $\mathrm{FSR}=v_{R}/2L$, which matches well with the vibrometer
measurement. We attribute the resonances between each pair of main
resonance dips to high-order modes along the thickness direction.
They have lower extinction ratios comparing to mian resonances since
the coupling between these modes and IDTs is relatively weak. Lorentzian
fitting is then applied to all mode. The resonance with highest quality
factor has a central frequency of $192.52\,\mathrm{MHz}$, with a
loaded Q of $1.37\times10^{4}$. Because all the modes are weakly
coupled to the external circuit, the intrinsic Q is nearly the same
with loaded Q. This quality factor is more than one order of magnitude
larger than the previous work on GaN SAW resonator, while the $f\times Q$
product ($\sim2.7\times10^{12}$) is improved by a factor of three
\cite{wang2015tapered} and becomes comparable to the suspended GaN
PnC resonators \cite{wang2014gan}. Nevertheless, this measured $f\times Q$
is still below the theoretical prediction \cite{rais2014gallium},
suggesting we have not yet reached the material limited qualty factor.
The quality factor can be further improved by further reducing the
radiation and the metal losses, for example, by increasing the cavity
length in $y$ direction or by eliminating the radiation loss at IDT-electrode
interfaces.

The resonance at $f_{0}=194.07\,\mathrm{MHz}$, on the other hand,
have the largest extinction ratio among all observed modes and is
selected to analyze the coupling efficiency between IDTs and SAW.
According to input-output fomula \cite{aspelmeyer2014cavity}, the
amplitude reflection of a cavity could be written as 
\begin{equation}
\mathcal{R}=\frac{\kappa_{int}-\kappa_{ext}-2i\varDelta}{\kappa_{int}+\kappa_{ext}-2i\varDelta}
\end{equation}
where $\varDelta=\omega-\omega_{0}$ is the frequency detuning from
the resonance frequency, and $\kappa_{int}$, $\kappa_{ext}$ are
the internal and external coupling coefficient. In our case, $\kappa_{int}=2\text{\ensuremath{\pi}}f_{0}/Q_{int}$
represents the intrinsic loss in cavity (including all the losses
except the loss to microwave channel), $\kappa_{ext}$ describes the
coupling strength between microwave and SAW through the IDTs. The
value of internal and external coupling coefficient could be extracted
from the Lorentzian fitting: $\kappa_{int}/2\pi=26.3\,\mathrm{kHz}$
, $\kappa_{ext}/2\pi=458\,\mathrm{Hz}$. If we consider the IDTs at
resonance frequency as a lumped element in transmission line, the
amplitude reflection is $\mathcal{R}=(Z_{\mathrm{eff}}-Z_{0})/(Z_{\mathrm{eff}}+Z_{0})$,
where $Z_{0}=50\mathrm{\,\Omega}$ is the characteristic source impedance
and $Z_{\mathrm{eff}}$ is the effective impedance of IDT and resonator.
If $\kappa_{int}$ and $Z_{0}$ are fixed, we have $\kappa_{ext}\propto1/Z_{\mathrm{eff}}$.
Here $Z_{\mathrm{eff}}$ is found to be $2.87\,\mathrm{k\Omega}$.
Since the IDTs should be effectively considered as parallel connected
component in circuit, the external coupling coefficient $\kappa_{ext}$
is propotional to the area of IDTs. By further increasing the area
of IDTs, a critical coupling between microwave and acoutic cavity
could be reached. In this work 5 pairs of IDT fingers are sufficient
to excite SAW.

\begin{figure}
\includegraphics[width=0.445\textwidth]{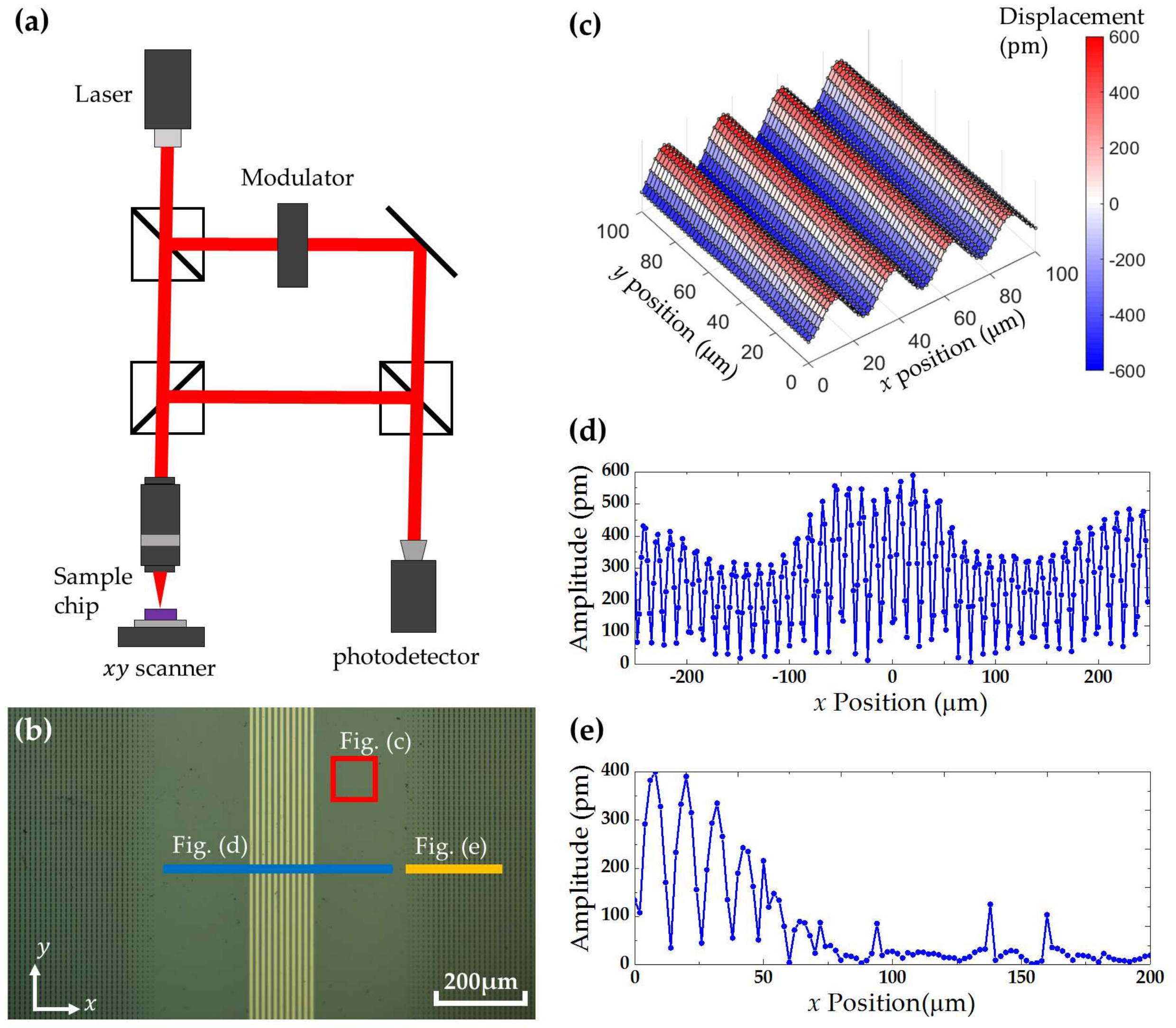}\caption{(a) Schematic of the optical heterodyne vibrometer used in imaging
acoustic resonators. (b) Optical micrograph of the device, where the
scan regimes are marked in red, blue and yellow, and the corresponding
measurement data are respectively shown in Fig. 3(c), (d) and (e).
(c) Out-of-plane displacement of the selected mode at 194.07 MHz acquired
by optical heterodyne vibrometer over a 100\textgreek{m}m\texttimes 100\textgreek{m}m
square inside the cavity. (d) Recorded out-of-plane vibration amplitude
along x-axis inside the cavity for the selected mode at 194.07 MHz.
(e) The vibration amplitude along x-axis in the PnC regime where SAW
attenuates rapidly ($\sim10a$) within the PnC mirror. The spikes
in the tail regime is due to laser reflection at hole edges and do
not correspond to a true displacement. }
\end{figure}

A custom-built optical heterodyne vibrometer \cite{zhen2017} is then
applied to assess and identify the modal profile of the resonator.
A simplified scheme of the optical heterodyne vibrometer is shown
in Fig. 3(a). The mode with central frequency of $194.07\,\mathrm{MHz}$
with largest extinction is chosen to be the imaged. Due to the limited
field of view of the vibrometer, we select three different scan locations
to assess the vibration mode: a square near the center of cavity (red
box in Fig. 3b), a line scan inside the cavity (blue line) and a line
scan extending outside the cavity into the PnC mirror along $x$ axis
(yellow line). The measurement results are shown in Figs. 3(c), (d)
and (e) respectively. From Figs. 3(c) and (d), we confirm that a well
confined Rayleigh mode with sinusoidal out-of-plane displacement is
obtained. Love mode have not been observed under the vibrometer, due
to the low excitation efficiency of IDTs for Love mode. And Fig. 3(e)
shows a smooth and fast attenuation of $z$ direction displacement
amplitude after the acoustic wave enters PnC regime, indicating that
100 periods of elliptical holes we use is redundant. A PnC mirror
with 20 periods of PnC holes would be sufficient to confine high Q
acoustic mode within the cavity, reducing the mirror length to less
than $220\,\mathrm{\mu m}$. 

\section{Conclusion}

In conclusion, we have designed and fabricated high-Q SAW Fabry-Perot
resonators at the frequency around $200\,\mathrm{MHz}$, with highest
$Q$-factor about $1.3\times10^{4}$. Through vibrometer measurement,
confined Rayleigh mode is directly visualized inside the cavity and
evanescent wave observed at the phononic crystal mirror. Our approach
can also be extended to other frequencies and may achieve even higher
$Q$, particularly at cryogenic temperatures or with other materials.
The demonstrated high-Q resonator is compatible with the sensing applications
(such as mass sensor \cite{luo2013new} and gyroscope \cite{jose2002surface})
and also is promising for the study of strong phonon-matter interaction,
especially when the device is advanced to operate at tens of GHz under
cryogenic condition.

\textbf{Acknowledgment}: We acknowledge useful discussions with M.-H.
Shen. This work is supported by DARPA/MTO\textquoteright s PRIGM:
AIMS program through a grant from SPAWAR (N66001-16-1-4026), the Laboratory
of Physical Sciences through a grant from Army Research Office (W911NF-14-1-0563),
an Air Force Office of Scientific Research (AFOSR) MURI grant (FA9550-15-1-
0029) and a NSF MRSEC grant (1119826). H.X.T. acknowledges support
from a Packard Fellowship in Science and Engineering. The authors
thankand Dr. Michael Rooks, Michael Power, James Agresta, and Christopher
Tillinghast for assistance in device fabrication.

\bibliographystyle{apsrev4-1}
\bibliography{Reference}

\end{document}